\documentclass[preprint,showpacs,preprintnumbers,amsmath,amssymb]{revtex4}

\usepackage{amsmath}
\usepackage{epsfig}
\usepackage{graphicx}

\begin{document}
\title{EMC effect in semi-inclusive deep-inelastic scattering process}

\author{Baogui Lu}
\affiliation{Department of Physics, Peking University, Beijing
100871, China}
\author{Bo-Qiang Ma}
\email{mabq@phy.pku.edu.cn} 
\affiliation{ CCAST (World Laboratory), P.O.~Box 8730, Beijing 100080, China\\
Department of Physics, Peking University, Beijing 100871,
China\footnote{Mailing address.}}

\begin{abstract}
By considering the $x$-dependence of $\pi^+$, $\pi^-$, $K^+$, $K^-$,
$\Lambda$, $\bar{\Lambda}$, $p$, $\bar{p}$ hadron productions in
charged lepton semi-inclusive deep inelastic scattering off nuclear
target
(using Fe as an example) and deuteron D target, 
we find that
$(\bar{\Lambda}^A/\Lambda^A)/(\bar{\Lambda}^D/\Lambda^D)$ and
$({\bar{p}}^A/{p}^A)/({\bar{p}}^A/p^A)$ are ideal to figure out the
nuclear sea content, which is predicted to be different by different
models accounting for the nuclear EMC effect.
\end{abstract}



\vfill

\maketitle

\section{\label{sec:level1}Introduction}

\begin{figure}
  \includegraphics[width=300pt]{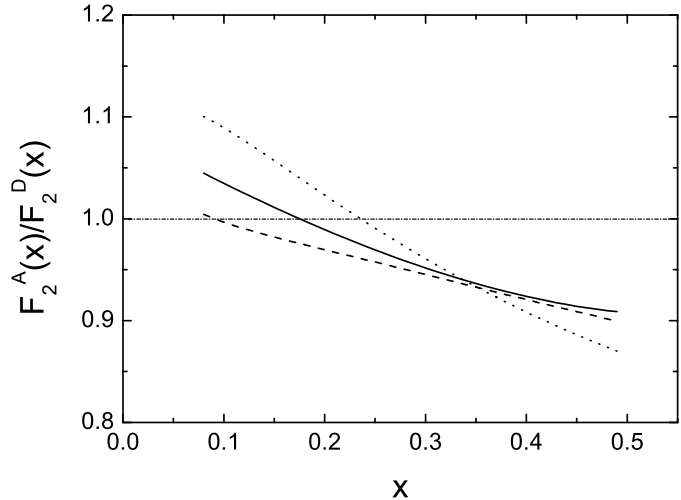} 
  \caption{\small {Results of $F_2^A(x)/F_2^D(x)$ in three models at $Q^2=5$~GeV$^2$. The solid, dashed and dotted
  curves
  are the results of the cluster
  model, the rescaling model and the pion excess
  model respectively. The target nuclei assumed here is Fe.}}\label{f2}
\end{figure}

\begin{figure}
  \includegraphics[width=350pt]{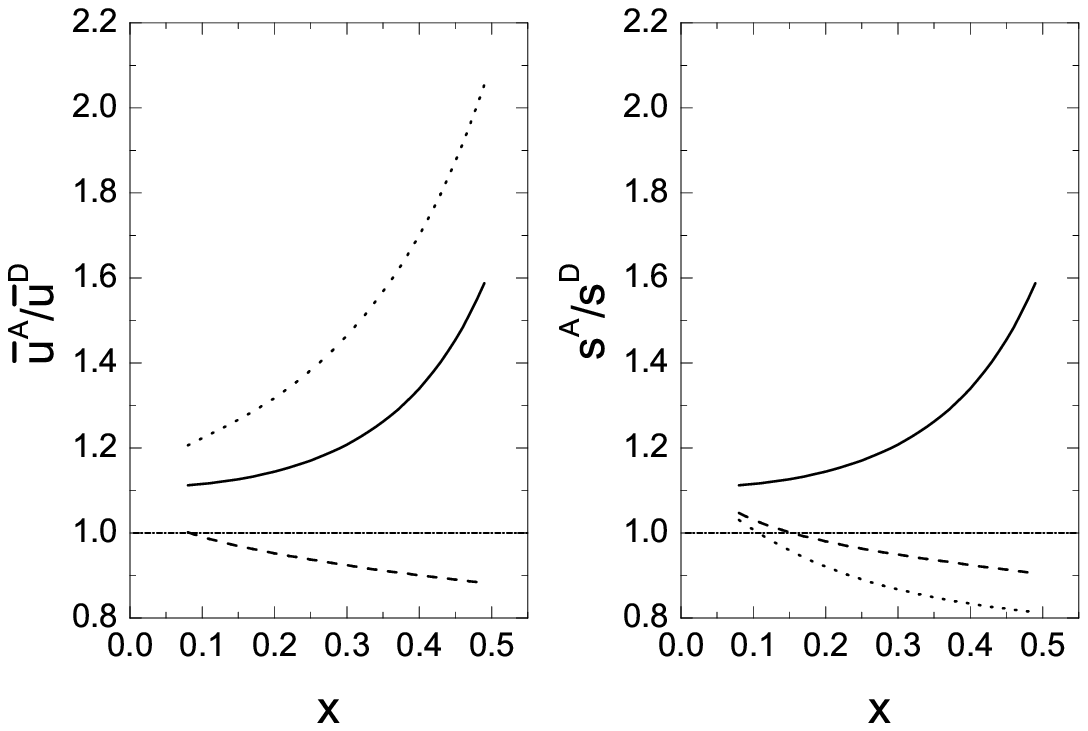} 
  \caption{\small {$\bar{u}$ and $s$ sea quark behaviors of various
  models.
  The solid, dashed and dotted curves correspond to the
  predictions of the
  cluster model, the rescaling model and the pion excess model respectively. The parton distribution is calculated at
  $Q^2=5$~GeV$^2$}. The target nuclei assumed here is Fe.}\label{sea}
\end{figure}

In 1982, the European Muon Collaboration (EMC) at CERN found that
the structure function ratio of bound nucleon to free nucleon, in
the form of $F^A_2(x,Q^2)/F^D_2(x,Q^2)$, is not consistent with the
expectation by assuming that a nuclei is composed by almost free
nucleons with Fermi motion correction taken into
account~\cite{EMCA,EMC}, and such phenomenon was confirmed by E139
collaboration at SLAC~{\cite{SLAC}}. This discovery, which is called
the nuclear EMC effect, has received extensive attention by the
nuclear and hadronic physics society. Many nuclear models, such as
the pion excess model~\cite{LEST,ME}, the quark cluster
model~\cite{Jaffe,Carlson,Vary} and the rescaling
model~\cite{Jaffe,FEC,RLJ,FECB,ONachtmann} {\it et al.}, have been
proposed to explain the data, and all these models can qualitatively
describe the data in the mediate $x$ region. The inclusive deep
inelastic scattering (DIS) data are expressed as
${F^A_2(x,Q^2)}/{F^D_2(x,Q^2)}$, which can be written in the naive
parton model as:
\begin{equation}
\frac{F^A_2(x,Q^2)}{F^D_2(x,Q^2)}
=\frac{\Sigma_i{e_i^2\left[q_i(x,Q^2,A)+\bar{q}_i(x,Q^2,A)\right]}}{\Sigma_i{e_i^2\left[q_i(x,Q^2)+\bar{q}_i(x,Q^2)\right]}},
\end{equation}
where $e_i$ denotes the charge of the partons with flavor $i$, and
$q(x,Q^2)$ is the parton distribution function of a nucleon.
Fig.~\ref{f2} shows the $F_2^A(x,Q^2)/F_2^D(x,Q^2)$ results of the
cluster model, the pion excess model and the rescaling model
respectively at $Q^2=5$~GeV$^2$ in the mediate $x$ region.
All these models, as can be seen from Fig.~\ref{f2}, predict similar
behavior of $F_2^A(x,Q^2)/F_2^D(x,Q^2)$ at mediate $x$ region.
However, the sea quark  of the nuclei is differently described by
the three models. In the cluster model, all sea quarks are enhanced.
In the pion excess model, the sea quarks $\bar{u}$ and $\bar{d}$ are
enhanced while the other quarks are reduced.  However, in the
rescaling model, all sea quarks are reduced in the nuclei compared
with those in the free nucleon (Fig.~\ref{sea}).

The Fermilab experiment 772~\cite{DMA} measured the dimuon yield in
Drell-Yan process induced by 800~GeV proton off various nucleus and
compared the data with the theoretical predictions of the three
models accounting for the EMC effect~\cite{DMA}. The data were
explained in Ref.~\cite{DMA} to favor a conclusion that the sea
quark in the nuclei is not enhanced, by neglecting the energy loss
effect of the incident quark, which is not precisely determined
yet~\cite{Mav,Mbj,Gtg}. To avoid the uncertainties concerning the
sea quarks in the nuclei by the dimuon yield in Drell-Yan process
solely, the sea content can be also measured in other experiments.
The purpose of this work is to show that the semi-inclusive hadron
productions in charged lepton deep inelastic scattering
are sensitive to the sea quark content of the nuclei. The pion,
kaon, proton and antiproton productions in the charged lepton
semi-inclusive deep inelastic scattering off nuclei have been
checked by the HERMES collaboration~\cite{Hermes}. However, while
the HERMES data, which are expressed as multiplicity ratios and also
with $z$-dependence, are convenient to study the modification of the
fragmentation functions in nuclear environment~\cite{Hermes}, they
are not ideal to provide much information about the sea quarks of
the nuclei. We will show that the $\bar{\Lambda}/{\Lambda}$
production, especially the $x$ dependent behavior, is ideal to
distinguish between different predictions on the sea content of the
nuclei in the different models of the EMC effect.

\section{Nuclear models and the sea quark distributions}

The semi-inclusive hadron productions in charged lepton deep
inelastic scattering can be related with the quark distribution
functions as:
\begin{equation}
 \frac{d^3\sigma^{h} }{dxdydz}=\frac{4\pi\alpha
s}{Q^4}(1+(1-y)^2)
\sum_ie_i^2\left[q_i(x,Q^2)D_{q_i}^{h}(z,Q^2)+\bar{q}_i(x,Q^2)D_{\bar{q}_i}^{h}(z,Q^2)
\right ],
\end{equation}
where $q_i(x,Q^2)$ is the parton distribution for quarks with
flavor $i$, and $D_{q_i}^h(z,Q^2)$ is the fragmentation function
of quark $q_i$ to hadron $h$. The formula is also applicable to
the nuclei, with the parton distributions and fragmentation
functions replaced by $q(x,Q^2){\rightarrow}q(x,Q^2,A)$ and
$D(z,Q^2){\rightarrow}D(z,Q^2,A)$ respectively. The inclusive
production itself can not offer enough information about the sea
quark enhancement, while the ratio
 \begin{equation}
    \frac{d\sigma^{h}_A/dx}{d{\sigma}^{h}_D/dx}=\frac{\int_{a}^{b}
    dz{\sum}_ie_i^2(q_i^A(x,Q^2)D_{q_i}^{h}(z,Q^2,A)+{\bar{q}}_i^A(x,Q^2)
    D_{{\bar{q}}_i}^{h}(z,Q^2,A))}{\int_{a}^{b}dz{\sum}_ie_i^2(q_i^D(x,Q^2)
    D_{q_i}^{h}(z,Q^2)+{\bar{q}}_i^D(x,Q^2)D_{{\bar{q}}_i}^{h}(z,Q^2))},
 \end{equation}
is useful to reveal the difference between the sea quark behavior
in the nuclei and that in the nucleon.

In the following, we will consider the ratios
$(d\sigma^{h}_A/dx)/(d{\sigma}^{h}_D/dx)$ for various hadrons. We
use the pion excess model, the rescaling model and the pion excess
model to calculate the sea quark content of the nuclei. For the sea
quark distributions in the deuteron D, we use the result offered by
the model itself~\cite{Carlson} for the cluster model, and for the
other two models we adopt the CTEQ5L parametrization~\cite{cteq} of
parton distributions for free nucleons by considering the isospin
symmetry between proton and neutron.

In the pion excess model, the quark distribution in the nuclei is
modified by the extra pions caused by the interaction between the
nucleons in nuclei~\cite{LEST}.
The quark distribution of nuclei is:
\begin{equation}
q_i^A(x)=\int_x^1{\frac{dy}{y}f_\pi^A(y)q_i^\pi(\frac{x}{y})}+\int_x^1{\frac{dy}{y}f_N^A(y)q_i^N(\frac{x}{y})},
\end{equation}
in which $q_i^\pi(x)$ and $q_i^N(x)$ are the parton distributions in
the free pion and in the free nucleon respectively, and $f_\pi(y)$
is the probability to find extra pions in the nuclei~\cite{ME}.
For simplicity, we adopt the parametrization in a toy
model~\cite{ELB}, in which the proton is supposed to be partially in
the nucleon-pion subsystem state and the parton distributions in the
nucleon and in the pion are assumed to be the same as those in the
free nucleon and in the free pion. Thus, the excess pion and the
nucleon probabilities per nucleon are given as~\cite{ELB}:
\begin{eqnarray}
  f_{\pi}^A(y) &=& \langle{n_{\pi}}\rangle\frac{\Gamma(a+b+2)}{\Gamma(a+1)\Gamma(b+1)}y^a(1-y)^b, \\
  f_N^A(z) &=&
  (1-\langle{n_{\pi}}\rangle)\delta(z-1)+f_{\pi}^A(1-z),
\end{eqnarray}
where $n_{\pi}=0.22$, $a=1$ and $b=3$. The CTEQ5L~\cite{cteq}
parametrization of the parton distribution of the nucleon
 and MRS~\cite{MRS} parametrization of the parton distribution of the pion are adopted to obtain
$F_2^A(x)$ and the parton distribution of the nuclei.

In the quark cluster model, six or more quark cluster is supposed to
exist in the nuclei to account for the EMC effect. For the sake of
simplicity, only six quark cluster is considered here. $q(x)$ in a
six-quark cluster can not be measured directly from experiment, but
Carlson and Havens~\cite{Carlson} estimated quark distribution q(x)
per nucleon in the six quark cluster based on QCD counting rules:
           \begin{eqnarray}
             v_6(x) &=& N_vz^{(-1/2)}(1-z)^{10}, \\
             \bar{u}_6(x) &=& (N_{sea}/4)z^{(-1)}(1-z)^{14},
           \end{eqnarray}
where $N_v=1.3875$ and $N_{sea}=0.2521$ are the coefficients to
warrant momentum conservation and the quark number. In the six quark
cluster, $x$, the variable defined by $Q^2/(2M_N\nu)$, equals to
$2z$ because $z$ is defined by $Q^2/(2M_6\nu)$~\cite{Carlson}.
Therefore, $q^A(x)$ and $F_2^A(x,Q^2)/F_2^N(x,Q^2)$ can be given as:
\begin{eqnarray}
  q^A(x) &=& (1-f)q^N(x)+fq^6(x), \\
  \frac{F_2^A(x,Q^2)}{F_2^N(x,Q^2)}&=&
  (1-f)+f\frac{F_2^6(x,Q^2)}{F_2^N(x,Q^2)},
\end{eqnarray}
where $f$ is the probability to find the six-quark cluster in the
nuclei and is adjusted to fit the inclusive deep inelastic
$e(\mu)$-$A$ scattering data. Its value is given as 0.30 by Carlson
and Havens~\cite{Carlson}.

For the rescaling model, the quark in the nuclear medium is
considered to have different confinement size compared with that of
the quark in the free nucleon. $q^A(x,Q^2)$ is related with
$q^N(x,Q^2)$~(the parton distribution in the free nucleon) by the
relation:
  \begin{equation}
q^A(x,Q^2)=q^N(x,\xi(Q^2)Q^2),
\end{equation}
where $\xi(Q^2)$ varies with $A$ and $Q^2$. $\xi$ equals to
$1.83$~\cite{FECB} while $Q^2=5$~GeV$^2$ and A=56~(Fe). For
$q^N(x,Q^2)$, the parton distribution per nucleon, CTEQ5L
parametrization~\cite{cteq} is adopted.

Given the above analysis, $F^A_2(x,Q^2)/F^D_2(x,Q^2)$ and sea
quark enhancement are checked in Fig.~\ref{f2} and Fig.~\ref{sea}.

\section{Fragmentation function and parton energy loss model}

Due to the non-perturbative nature of the fragmentation process, the
fragmentation function can not be calculated from first principle,
thus models are used to obtain the fragmentation function.
Experimentally, the process of
 $e^+ + e^- \rightarrow h+X$ can offer much information about the fragmentation~\cite{pdg}.

Based on the experimental data and theoretical analysis,
Kretzer~\cite{Kre} gave a parametrization of the fragmentation in
the form:
       \begin{equation}
            D_a^h(x,Q_0^2)=Nx^\alpha(1-x)^\beta,
       \end{equation}
where $N$, $\alpha$ and $\beta$ are the constants chosen to fit the
experimental data. For specified hadron, $\alpha$, which determines
the low $z$ region behavior of the fragmentation function, is the
same for all light flavor quarks while $\beta$, related with the
high $z$ region behavior of the fragmentation function, is different
for various quarks. Therefore, in low $z$ region, all quark
fragmentation functions have the same shape, and in the large $z$
region, the favored quark fragmentation function is larger than the
unfavored quark fragmentation function. In addition, because of the
strange dominance at large $x$ for $s$ in $K^-$, the fragmentation
function such as $s\rightarrow K^-$ is larger than that of
$\bar{u}\rightarrow K^-$.

As for $\Lambda,\bar{\Lambda},p,\bar{p}$, there exists a
phenomenological parametrization~\cite{Ma} of their fragmentation
functions based on the assumption that the fragmentation function of
quark $q$ to hadron $h$ is proportional to the $q$ quark
distribution in the hadron $h$:
\begin{equation}
 D_q^h(z)\propto q^h(z).
\end{equation}
In general, the fragmentation functions can be written
as:
\begin{eqnarray}
D_V^h(z) &=& C_V(z)z^\alpha q_V^h(z), \\
D_S^h(z) &=& C_S(z)z^\alpha q_S^h(z),
\end{eqnarray}
      where $D_S^h(z)$ means unfavored fragmentation function.
      There are three options for the favored and unfavored quark fragmentations:
      (1). $C_V=1$ and $C_S=0$ for $\alpha=0$; (2). $C_V=1$ and $C_S=1$ for
      $\alpha=0.5$; (3). $C_V=1$ and $C_S=3$ for
      $\alpha=1$.
      The parton distributions of $\Lambda$ and $\bar{\Lambda}$
      are
      essential to get the fragmentation functions of quark to $\Lambda$ and
      $\bar{\Lambda}$. Based on the fact that there is no direct parton distribution of  $\Lambda$ and
      $\bar{\Lambda}$, SU(3) symmetry
       between the proton and the $\Lambda$ is adopted to get the parton distribution in
       $\Lambda$~\cite{Ma}. Although there are three models, SU(3)
       symmetry model, quark-diquark model and pQCD based analysis,
       the difference of the fragmentation functions will not affect the qualitative predictions
       by the
       fact that only the flavor structure of
       the parton distributions of the proton are different in the three
       models in the high $x$ region~\cite{MaB}.


The common feature of the above parameterizations of fragmentation
functions is that the favored quark fragmentation is larger than the
unfavored quark fragmentation in the large $z$ region. Thus the
favored quark fragmentation process is able to obtain sea quark
information of the nuclei if the produced hadron is from the favored
fragmentation of sea quarks in the nuclei. Then the hadron events of
produced particles in the large $z$ region can be chosen to get the
$x$ dependence of the production ratio, in which the unfavored
fragmentation contribution from the valence quark can be largely
suppressed.

Fragmentation function in the nuclei is important for producing the
hadron production from the nuclei. HERMES collaboration has measured
the hadron production from the nuclei and found that the production
is reduced compared with that from the free nucleon~\cite{Hermes},
and many effects such as nuclear absorption~\cite{bialas}, parton
energy loss~\cite{XNW,XG}, gluon bremsstrahlung~\cite{kopeli} and
partial deconfinement~\cite{RLJ,FECB,ONachtmann,partial} have been
developed to account for the data. In this paper, the parton energy
loss model is adopted to get the fragmentation function in the
nuclei.



In Refs.~\cite{XNW,XG}, the modification of the fragmentation
function is caused by the interaction between the hard quark and the
debris of the nuclei. Given that the original parton energy loss
model is complicated to apply in the real process, an effective
model suggested in Ref.~\cite{Xwn} is used in Ref.~\cite{Fran} to
get the modification of the fragmentation function. In effective
parton energy loss model, modified fragmentation function is
expressed in the form:
     \begin{equation}
        zD^h_q(z,Q^2,A)=\int_0^{(\nu-E_h)}d\epsilon{D(\epsilon,\nu)}z^*D^h_q(z^*,Q^2),
     \end{equation}
where $E_h=\mu-\epsilon$. $E_h$ is the measured hadron energy and
$\epsilon$ is the energy loss of the hard quark going through the
nuclei. $z^*$ is the rescaled momentum fraction caused by the quark
energy shift in presence of QCD medium:
     \begin{equation}
        z^*=\frac{E_h}{1-\left( \frac{\epsilon}{\nu}\right ) }.
     \end{equation}
 $D(\epsilon,Q^2)$, the probability for a quark with energy $E=\nu$ to lose energy $\epsilon$,
  is parameterized by Arleo~\cite{Arleo}:
      \begin{equation}
        D(\epsilon)=\frac{1}{\sqrt{2\pi}\sigma\epsilon}{\mathrm{exp}}[-\frac{(\mathrm{log}(\epsilon/\omega_c)-\mu)^2}{2\sigma^2}],
      \end{equation}
where $\mu$, $\sigma$ are two parameters with $\sigma=0.73$ and
$\mu=-1.5$, as the energy of the quark, which has absorbed the
virtual photon, is much higher than the energy loss when it passes
through the nuclei environment. And $\omega_c$ is the relevant scale
of the typical gluon energy and denotes the energy loss scale of the
hard quark,
   \begin{equation}
        \omega_c=\frac{1}{2}{\hat{q}}L^2.
   \end{equation}
Although $\hat{q}$ is not precisely determined yet, in the next
section we will show that the result is not sensitive on $\hat{q}$
when $Q^2$ is large enough. Here we set  $\hat q=0.72\ {\rm
GeV}/{\rm fm}^2$ and $L=3/4R$, where $R$ is the nuclear
radius~\cite{Fran}.

From another point of view, the modification of the fragmentation
function in the parton energy loss model is from assumption. Its
conformation with the theoretical framework of factorization and
renormalization is not fully justified, as the definition of the
fragmentation functions are vacuum matrix elements with no relation
to the target material. As we will find, the qualitative conclusion
of our paper on the ratio
$(\bar{\Lambda}^A/\Lambda^A)/(\bar{\Lambda}^D/\Lambda^D)$ will not
be influenced by including the modification of the fragmentation
function in the nuclear environment. In order to get rid of the
nuclear absorption or the energy loss process, it would be fine to
consider hadron production at larger energy (say larger than
20~GeV), hence at higher virtuality, where we know from the present
HERMES data that these two effects prove negligible.

\section{Results}

\begin{figure}
  \includegraphics[width=350pt]{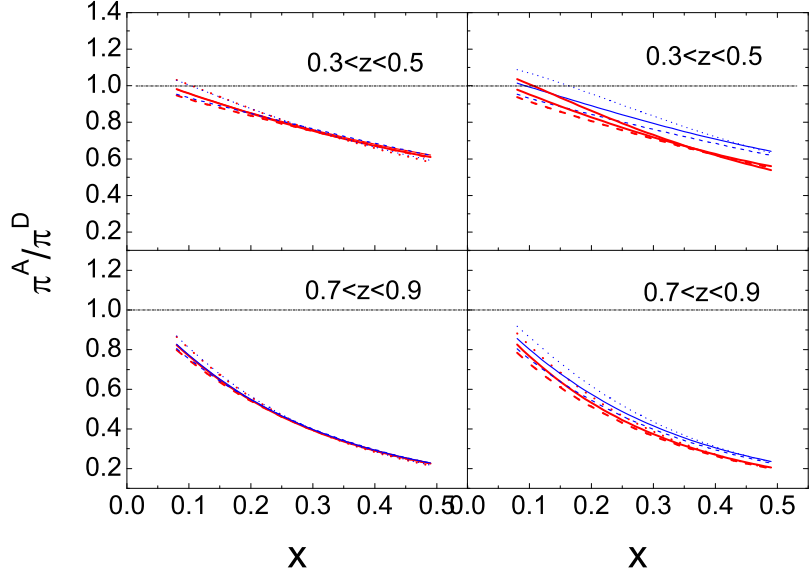} 
  \caption{\small {(Color online)The production ratios of $\pi^+,\pi^-$ at $Q^2=5$~GeV$^2$ in the region $0.7<z<0.9$ and
  $0.3<z<0.5$.
  The left two figures
  are the results of $\pi^+$ and the right are $\pi^-$. The upper two figures
  are in low $z$ region and the lower two are in high $z$ region. The solid, dashed and dotted curves denote
  the results of the cluster
  model, the rescaling model and the pion excess
  model respectively. The thick curves denote the production with all quark fragmentation and the thin
  curves correspond to the results with only favored quark fragmentation.
  The left two are $\pi^{+A}/\pi^{+D}$, and the right two are $\pi^{-A}/\pi^{-D}$
   The target nuclei assumed here is Fe.}}\label{pion}
\end{figure}

\begin{figure}
  \includegraphics[width=350pt]{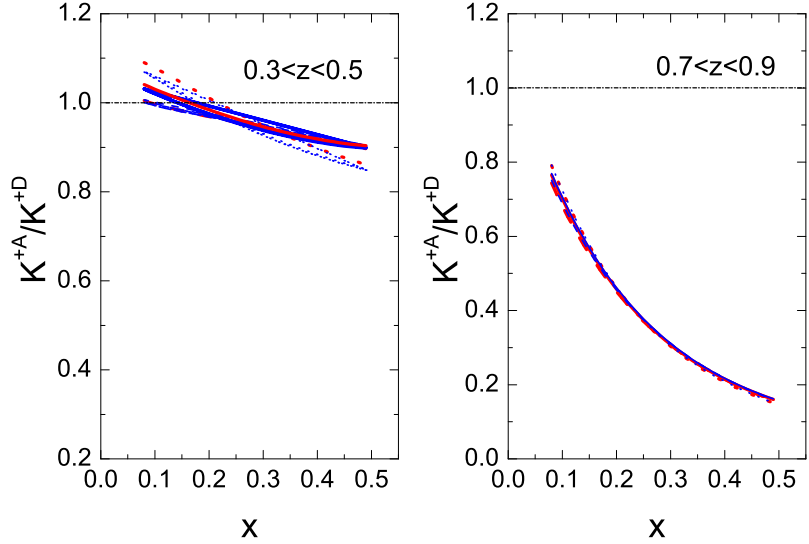} 
  \caption{\small {(Color online)The production ratio of $K^+$ at $Q^2=5$~GeV$^2$ in the region $0.7<z<0.9$ and $0.3<z<0.5$. The
left is the result in the low $z$ region and the right is the result
in the high $z$ region. The solid, dashed and dotted curves are the
predictions of the cluster model, the rescaling model and the pion
excess model respectively. The thick curves denote the production
with all quark fragmentation and the thin
  curves correspond to the results when only favored quark fragmentation. The target nuclei assumed here is Fe.}}\label{kplus}
\end{figure}

\begin{figure}
  \includegraphics[width=300pt]{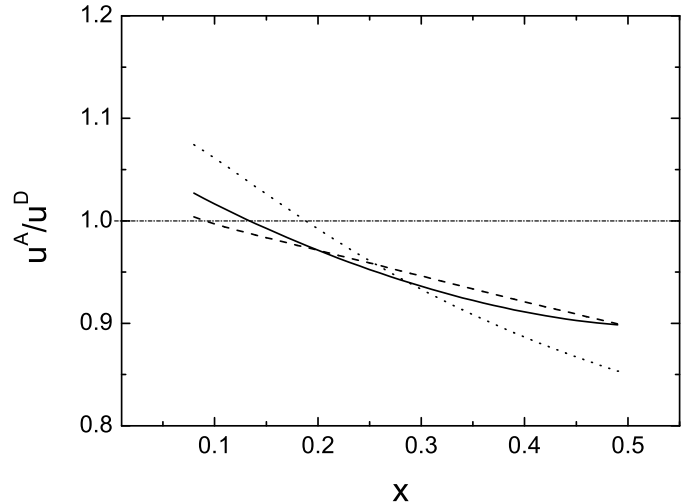} 
  \caption{\small{ The ratio $u^A(x)/u^D(x)$ of the three models at $Q^2$=5~GeV$^2$. The solid, dashed and dotted curves
  are the predictions of the cluster model, the rescaling model and the pion excess
model respectively. The target nuclei assumed here is
Fe.}}\label{uv}
\end{figure}

Fig.~\ref{pion} and Fig.~\ref{kplus} present the results of
$\pi^+(u\bar{d})$, $\pi^-(\bar{u}d)$ and $K^+$ respectively. They
show that there is no large difference between various model
predictions in the large $z$ region and in the low $z$ region, no
matter by including the favored quark fragmentation process only or
by including all favored and unfavored fragmentation processes. The
reason is that $\pi^+(u\bar{d})$, $\pi^-(\bar{u}d)$ and $K^+$ are
contributed mainly by the favored fragmentation processes
$u\rightarrow \pi^+$, $d\rightarrow \pi^-$ and $u\rightarrow K^+$
because that the valence quarks $u$ and $d$ are predominant over the
sea quarks in the mediate $x$ region and that production of those
hadrons are dominated by the behavior of the valence quark in the
nuclei (Fig.~\ref{uv}).

\begin{figure}
  \includegraphics[width=350pt]{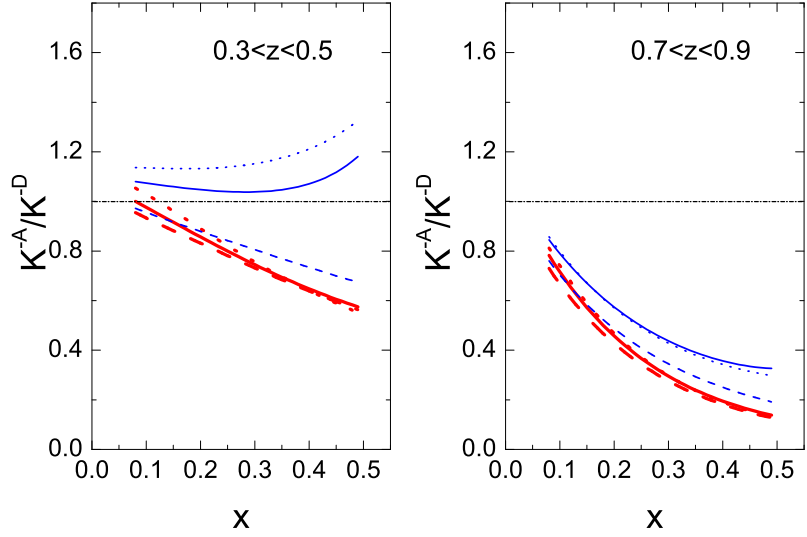} 
  \caption{\small {(Color online)The production ratio of $K^-$ at $Q^2=5$~GeV$^2$. The solid, dashed and dotted curves are the
predictions of the cluster model, the rescaling model and the pion
excess model respectively. The left figure is the result in the
region $0.3<z<0.5$ and the right is the result in the region
$0.7<z<0.9$. The thick curves denote the total results including all
quark fragmentation processes and the thin curves correspond to the
results when only favored quark fragmentation processes. The target
nuclei assumed here is Fe.}}\label{kminus}
\end{figure}

\begin{figure}
  \includegraphics[width=350pt]{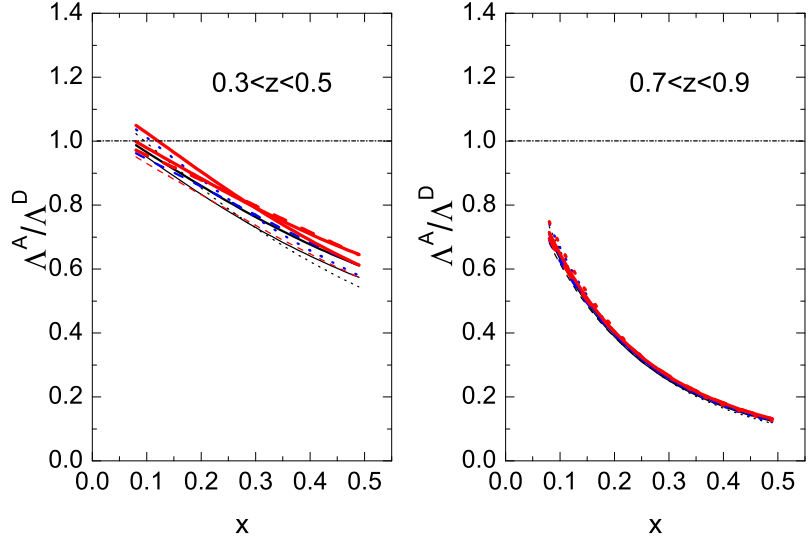} 
  \caption{\small {(Color online) The ratio $\Lambda^A/\Lambda^D$ at $Q^2=5$~GeV$^2$. The solid, dashed and dotted curves are the
  predictions of the cluster model, the rescaling model and the pion excess model respectively.
  The left figure is the result in the region $0.3<z<0.5$ and the right is the result in the region $0.7<z<0.9$,
  with the thin, normal and thick curves corresponding to the three options of the unfavored
  fragmentation (1), (2), (3) respectively. The target nuclei assumed here is Fe. }}\label{lambda}
\end{figure}

\begin{figure}
  \includegraphics[width=400pt]{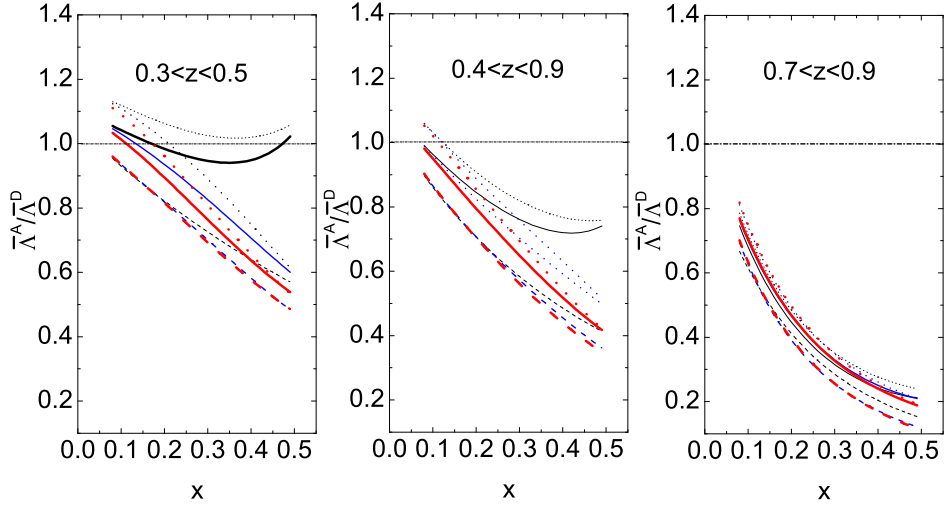} 
  \caption{\small {(Color online)The production ratio of $\bar{\Lambda}^A/\bar{\Lambda}^D$ at $Q^2=5$~GeV$^2$.
   The solid, dashed and dotted curves are the
  predictions of the cluster model, the rescaling model and the pion excess model respectively.
  The left figure is the result in the region $0.3<z<0.5$, the middle is the result in the region $0.4<z<0.9$
  and the right is the result in the region $0.7<z<0.9$.
  The thin, normal and thick curves correspond to the three options of the unfavored fragmentation
  (1), (2), (3) respectively. The target nuclei assumed here is Fe.}}\label{lambdabar}
\end{figure}

\begin{figure}
  \includegraphics[width=350pt]{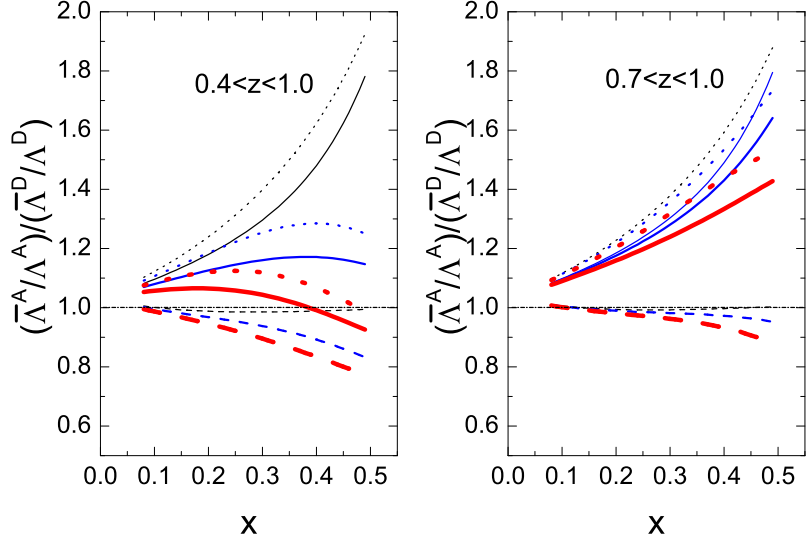} 
  \caption{\small {(Color online) The production ratio of $(\bar{\Lambda}^A/\Lambda^A)/(\bar{\Lambda}^D/\Lambda^D)$ at $Q^2=5$~GeV$^2$,
calculated in $0.7<z<1.0$ and $0.4<z<1.0$. The solid, dashed and
dotted curves are the predictions of the cluster model, the
rescaling model and the pion excess model respectively. The thin,
normal and thick curves correspond to the three options of the
  fragmentation (1), (2), (3) respectively. The target nuclei assumed here is Fe.}}\label{ratio}
\end{figure}

\begin{figure}
  \includegraphics[width=350pt]{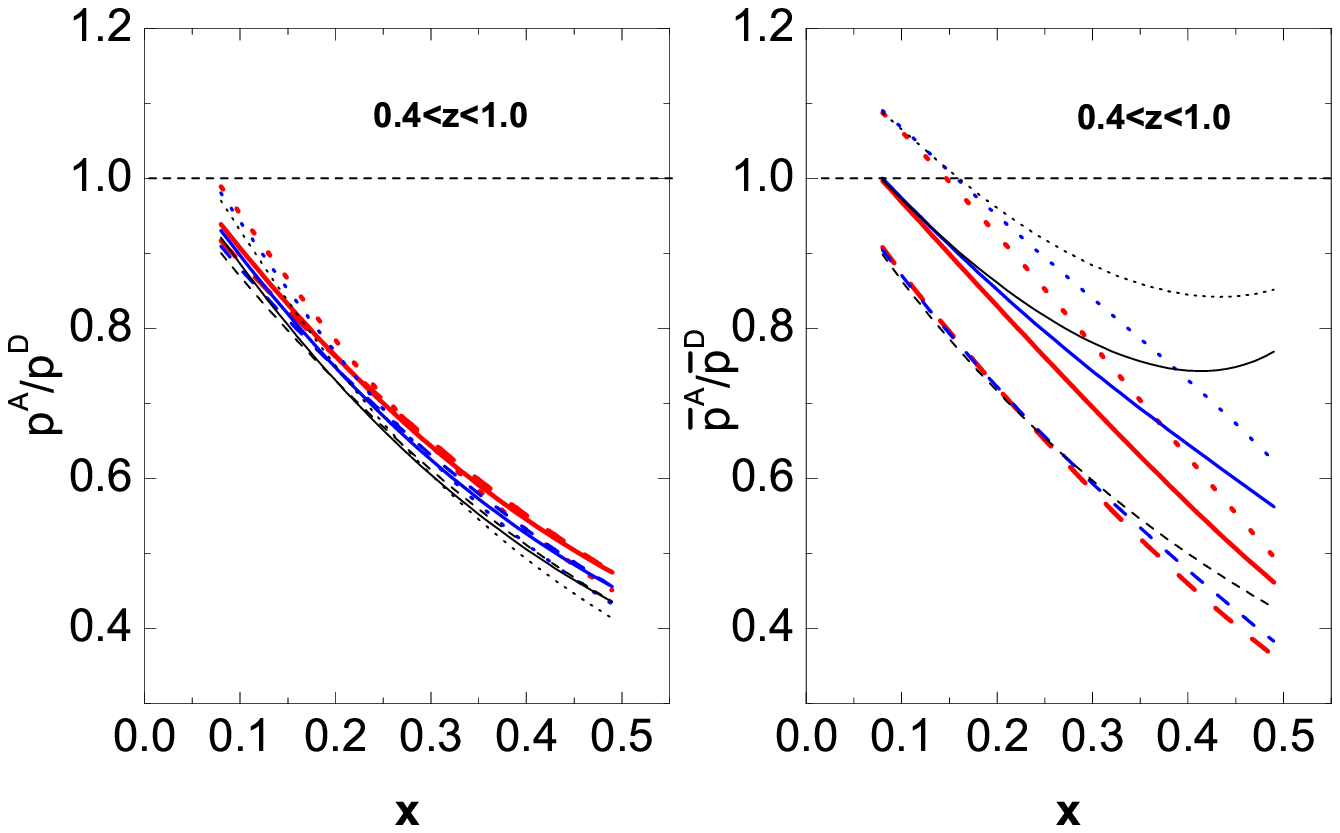} 
  \caption{\small {(Color online) The ratio $\bar{p}^A/\bar{p}^D$(right) and $p^A/p^D$(left) at $Q^2=5$~GeV$^2$. The solid, dashed and dotted curves are the
  predictions of the cluster model, the rescaling model and the pion excess model respectively.
  Thin, normal and thick curves correspond to the three options of the unfavored
  fragmentation (1), (2), (3) respectively. The target nuclei assumed here is Fe. }}\label{antiproton}
\end{figure}

\begin{figure}
  \includegraphics[width=350pt]{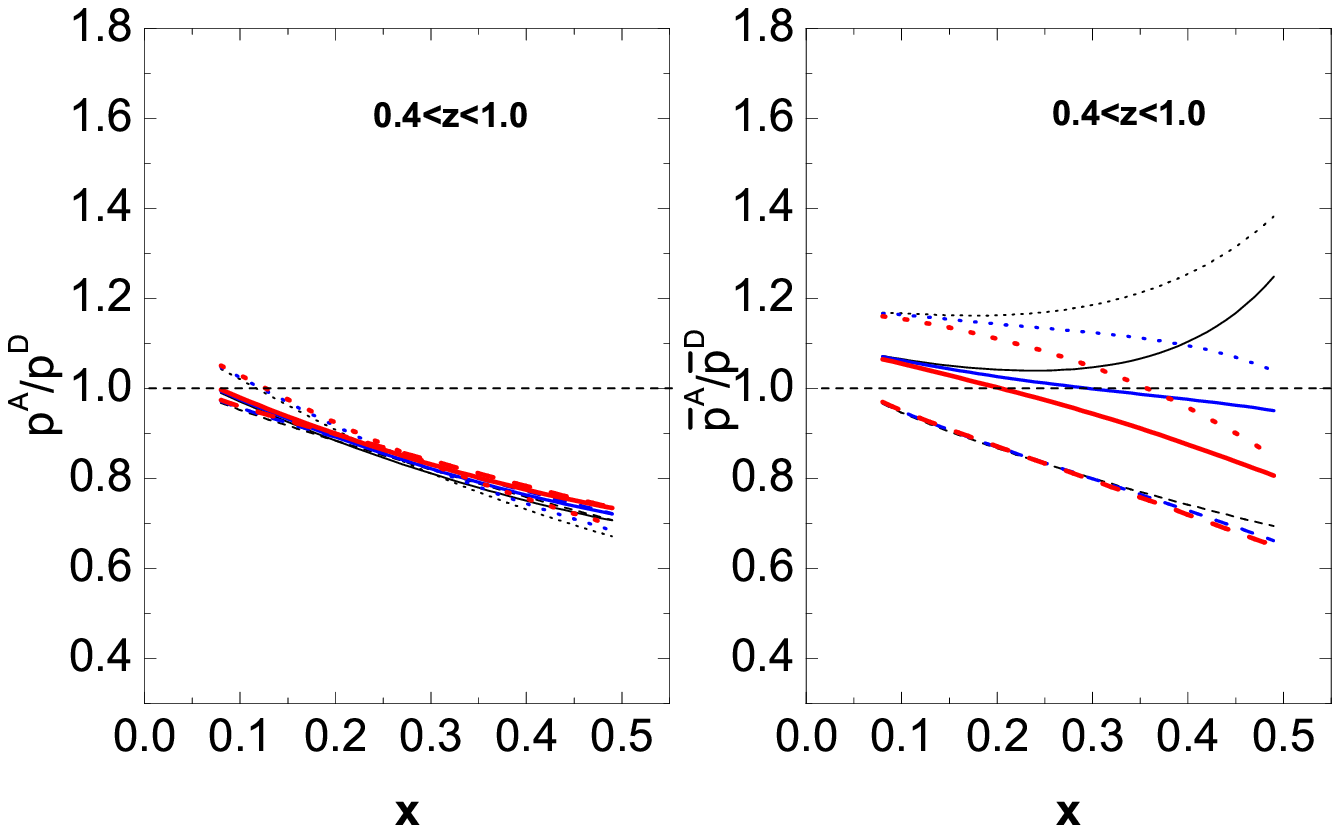} 
  \caption{\small {(Color online) The ratio $\bar{p}^A/\bar{p}^D$(right) and $p^A/p^D$(left) at $Q^2=5$~GeV$^2$ with $\hat q=0.25\ {\rm GeV}/{\rm fm}^2$. The solid, dashed and dotted curves are the
  predictions of the cluster model, the rescaling model and the pion excess model respectively.
  Thin, normal and thick curves correspond to the three options of the unfavored
  fragmentation (1), (2), (3) respectively. The target nuclei assumed here is Fe. }}\label{antiproton25}
\end{figure}

\begin{figure}
  \includegraphics[width=350pt]{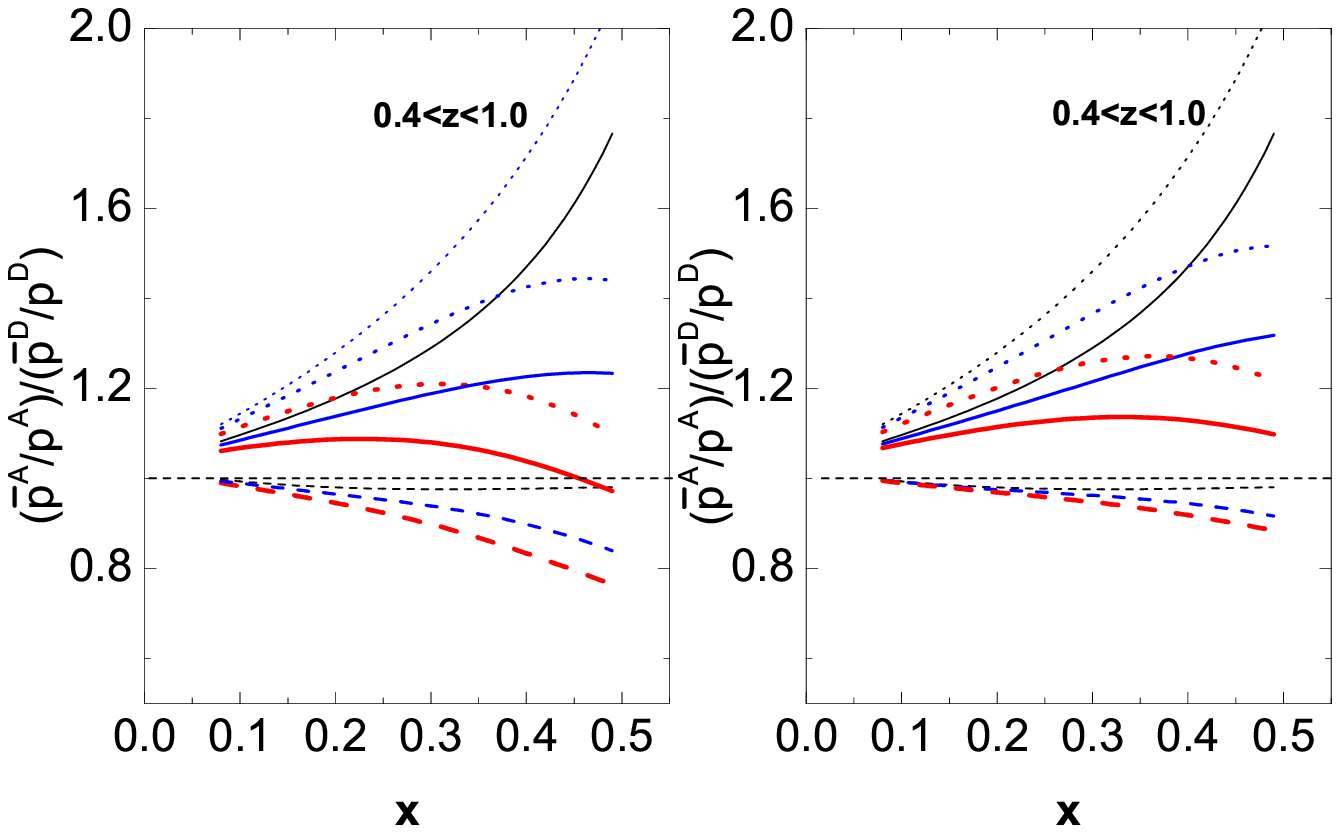} 
  \caption{\small {(Color online) The ratio $({\bar{p}}^A/{p}^A)/({\bar{p}}^A/p^A)$ at $Q^2=5$~GeV$^2$ with $\hat q=0.72\ {\rm GeV}/{\rm fm}^2$ (left) and $\hat q=0.25\ {\rm GeV}/{\rm fm}^2$(right). The solid, dashed and dotted curves are the
  predictions of the cluster model, the rescaling model and the pion excess model respectively.
  Thin, normal and thick curves correspond to the three options of the unfavored
  fragmentation (1), (2), (3) respectively. The target nuclei assumed here is Fe. }}\label{ratiop}
\end{figure}

\begin{figure}
  \includegraphics[width=350pt]{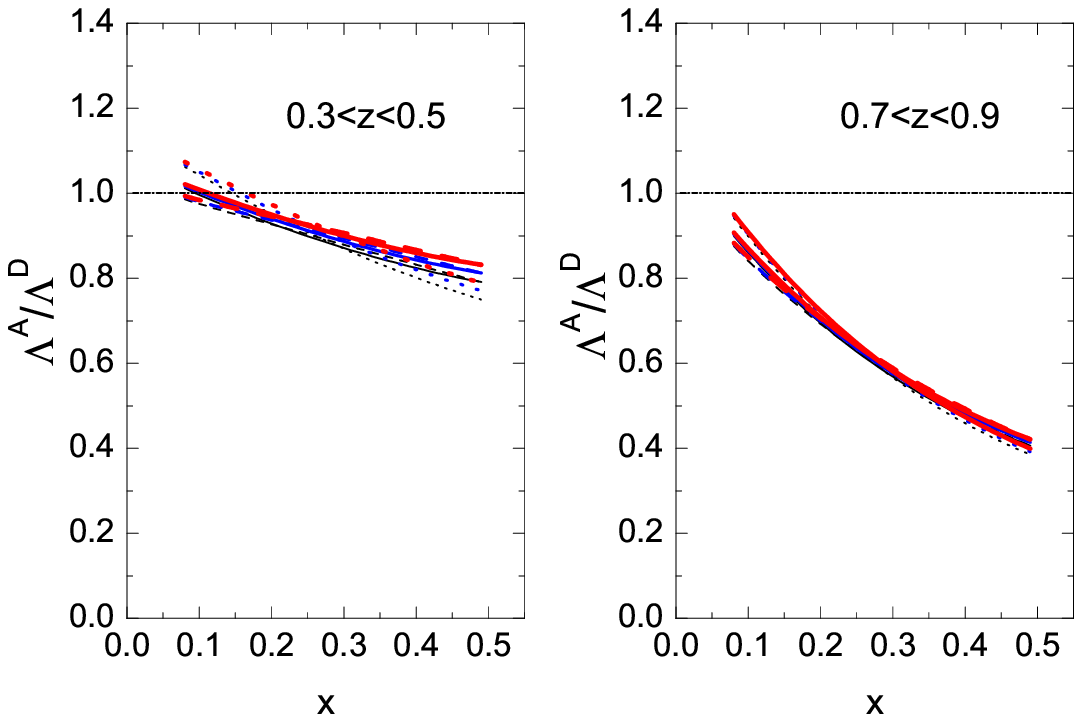} 
  \caption{\small {(Color online) The ratio $\Lambda^A/\Lambda^D$ at $Q^2=5$~GeV$^2$ with $\hat q=0.25\ {\rm GeV}/{\rm fm}^2$. The solid, dashed and dotted curves are the
  predictions of the cluster model, the rescaling model and the pion excess model respectively.
  The left figure is the result in the region $0.3<z<0.5$ and the right is the result in the region $0.7<z<0.9$,
  with the thin, normal and thick curves corresponding to the three options of the unfavored
  fragmentation (1), (2), (3) respectively. The target nuclei assumed here is Fe. }}\label{lambda25}
\end{figure}

\begin{figure}
  \includegraphics[width=400pt]{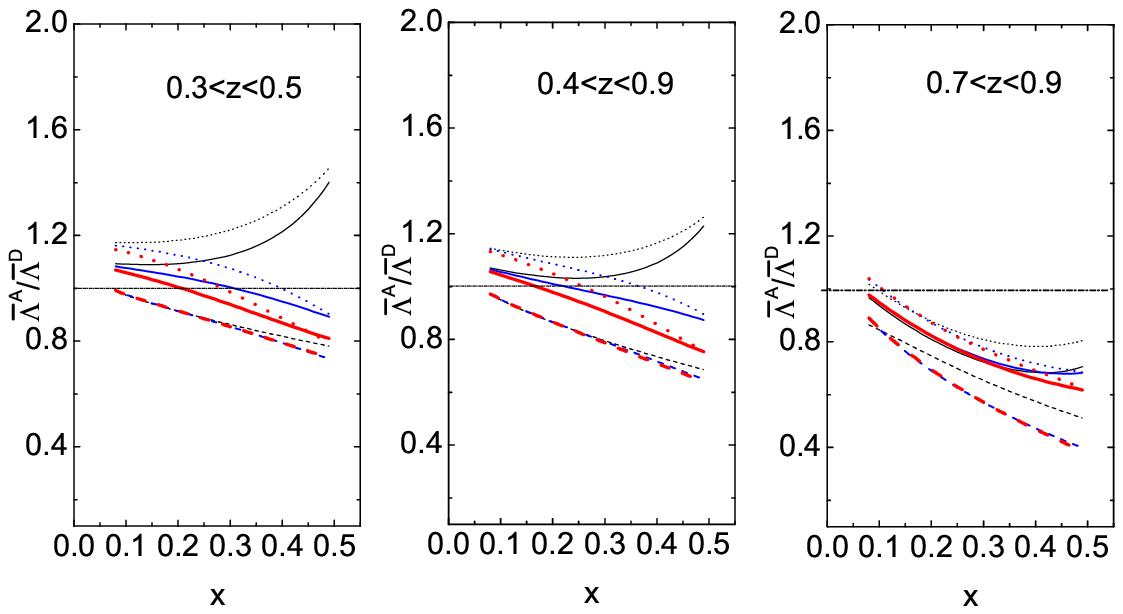} 
  \caption{\small {(Color online)The production ratio of $\bar{\Lambda}^A/\bar{\Lambda}^D$ at $Q^2=5$~GeV$^2$  with $\hat q=0.25\ {\rm GeV}/{\rm fm}^2$.
   The solid, dashed and dotted curves are the
  predictions of the cluster model, the rescaling model and the pion excess model respectively.
  The left figure is the result in the region $0.3<z<0.5$, the middle is the result in the region $0.4<z<0.9$
  and the right is the result in the region $0.7<z<0.9$.
  The thin, normal and thick curves correspond to the three options of the unfavored fragmentation
  (1), (2), (3) respectively. The target nuclei assumed here is Fe.}}\label{lambdabar25}
\end{figure}

\begin{figure}
  \includegraphics[width=350pt]{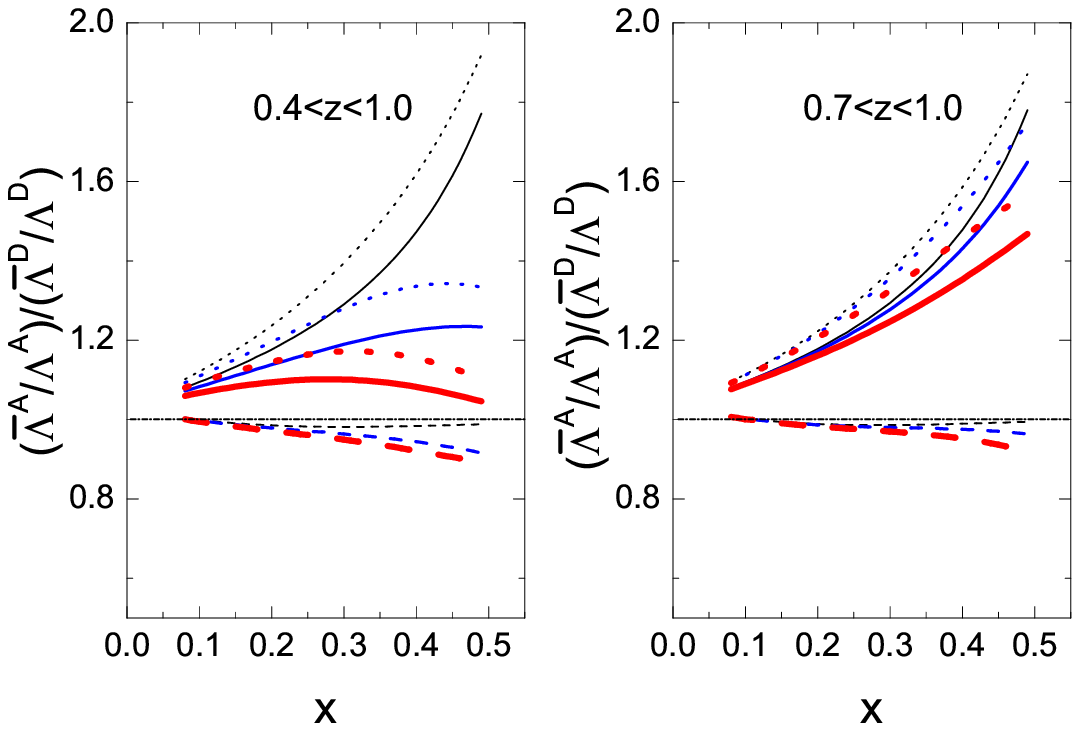} 
  \caption{\small {(Color online)The production ratio of $(\bar{\Lambda}^A/\Lambda^A)/(\bar{\Lambda}^D/\Lambda^D)$ at $Q^2=5$~GeV$^2$  with $\hat q=0.25\ {\rm GeV}/{\rm fm}^2$,
calculated in $0.7<z<1.0$ and $0.4<z<1.0$. The solid, dashed and
dotted curves are the predictions of the cluster model, the
rescaling model and the pion excess model respectively. The thin,
normal and thick curves correspond to the three options of the
  fragmentation (1), (2), (3) respectively. The target nuclei assumed here is Fe.}}\label{ratio25}
\end{figure}


When focused on $K^-$, the result with only favored quark
fragmentation functions is different from that when all
fragmentation processes are considered (Fig.~\ref{kminus}), which
indicates that the unfavored quark fragmentation function, $u$ and
$d$ to $K^-$, can not be neglected both at low and high $z$ region
for the predominance of the valence quarks in the $x$ region we
considered. In the high $z$ region, due to parton energy loss, the
$K^-$ is largely suppressed and we can hardly see any difference in
the three model predictions with all fragmentation processes being
considered.

$\Lambda$ and $\bar{\Lambda}$ production ratios in different $z$
regions are examined and the results are plotted in
Fig.~\ref{lambda} and Fig.~\ref{lambdabar}. For the same reason as
$K^+$, three models predict almost the same $x$-dependence of
$\Lambda^A/\Lambda^D$.  From Fig.~\ref{lambdabar}, difference are
generated among various model predictions on the $x$-dependence of
$\bar{\Lambda}^A/\bar{\Lambda}^D$, and the these difference are not
sensitive to the three options of the fragmentation
functions~\cite{Ma}. The reason is that the dominant production of
$\bar{\Lambda}$ is through the favored fragmentation of anti-quarks
inside the targets, so that the $x$-dependence of the production
ratio is sensitive to the sea quark behaviors of the nuclei. But in
the large $z$ region, due to the parton energy loss effect, the
large difference between three models do not manifest themselves
significantly as expected. The reason is, at large $z$ region, the
fragmentation function is largely modified by the parton energy
loss. And such a phenomenon also happens on $\Lambda^A/\Lambda^D$.
Fig.~\ref{lambdabar} shows that $\bar{\Lambda}^A/\bar{\Lambda}^D$ is
not an ideal variable to figure out the nuclear sea quark content by
including the modification of the fragmentation function in the
nuclear environment. Fortunately, largely difference between three
models appears for the quantity
$(\bar{\Lambda}^A/\Lambda^A)/(\bar{\Lambda}^D/\Lambda^D)$, which is
more accessible in experiment than
$\bar{\Lambda}^A/\bar{\Lambda}^D$. When the integral upper limit is
fixed with unity and the lower limit varies from 0.4 to 0.7,
$(\bar{\Lambda}^A/\Lambda^A)/(\bar{\Lambda}^D/\Lambda^D)$ is still
model dependent (Fig.~\ref{ratio}). Thus it is plausible to conclude
that $(\bar{\Lambda}^A/\Lambda^A)/(\bar{\Lambda}^D/\Lambda^D)$ can
offer information about the sea content of the nuclei.

Similar to $\bar{\Lambda}^A/\bar{ \Lambda}^D$, the antiproton and
proton production ratio $({\bar{p}}^A)/({\bar{p}}^D)$ can not offer
much information while ${\bar{p}}^A/{p}^A/{\bar{p}}^A/p^A$ do
generate large difference with different nuclear
model~(Figs.~\ref{antiproton}-\ref{ratiop}). So, the ratio
$({\bar{p}}^A/{p}^A)/({\bar{p}}^A/p^A)$ is another choice to check
the sea content of the nuclei in experiment. Attention should be
paid to extract possible background contribution as a large number
of protons and antiprotons might be produced from the decays of
other baryons.


$\hat{q}$ is a sensitive parameter that could largely affect the
modification of fragmentation function and is not determined clearly
yet. We also calculate $\Lambda^{A}/\Lambda^{D}$,
    $\bar{\Lambda}^A/\bar{\Lambda}^D$ and $(\bar{\Lambda}^A/\Lambda^A)/(\bar{\Lambda}^D/\Lambda^D)$
    when  $\hat q=0.25\ {\rm GeV}/{\rm fm}^2$ (Figs.~\ref{lambda25}-\ref{ratio25}). From the figures we can conclude that
    $\Lambda^{A}/\Lambda^{D}$ and
    $\bar{\Lambda}^A/\bar{\Lambda}^D$ are largely affected by
    different $\hat{q}$, while
    $(\bar{\Lambda}^A/\Lambda^A)/(\bar{\Lambda}^D/\Lambda^D)$ is
    almost $\hat{q}$ independent. And
    $({\bar{p}}^A/{p}^A)/({\bar{p}}^A/p^A)$ also has such
    properties(Figs.~\ref{antiproton25}-\ref{ratiop}).
So, the ratio
$(\bar{\Lambda}^A/\Lambda^A)/(\bar{\Lambda}^D/\Lambda^D)$ and
$({\bar{p}}^A/{p}^A)/({\bar{p}}^A/p^A)$ are not sensitive to
$\hat{q}$.

Besides, we should mention that the fixed order calculation used
here is not appropriate to describe hadron production at high $z$
where large logarithms need to be resumed. Therefore we should
consider the results here as qualitative predictions rather than
quantitative ones. For more convinced quantitative predictions, we
would need better constrained fits of the nuclear parton
distributions, rather than the earlier EMC model results adopted in
this paper.

\section{Summary}

In this paper, we adopted three models of the nuclear EMC effect:
the cluster model, the rescaling model and the pion excess model, to
calculate their predictions on the hadron production ratio in
charged lepton semi-inclusive deep inelastic scattering off nuclei
in the large $z$ region. Our purpose is to find hadrons which are
produced mainly from the sea quarks of nucleus, so that we can
distinguish between different predictions on the sea content of the
nuclei. For completeness, we considered the production ratios of
$\pi^{+A}/\pi^{+D}$, ${\pi^{-A}/\pi^{-D}}$, $K^{+A}/K^{+D}$,
    $K^{-A}/K^{-D}$, $\Lambda^{A}/\Lambda^{D}$,
    $\bar{\Lambda}^A/\bar{\Lambda}^D$,$(\bar{\Lambda}^A/\Lambda^A)/(\bar{\Lambda}^D/\Lambda^D)$, $({\bar{p}}^A/{p}^A)/({\bar{p}}^A/p^A)$
    and found that the ratios of
    $(\bar{\Lambda}^A/\Lambda^A)/(\bar{\Lambda}^D/\Lambda^D)$ and $({\bar{p}}^A/{p}^A)/({\bar{p}}^A/p^A)$ are ideal to figure
    out the sea content of the nuclei.

More significantly,
$(\bar{\Lambda}^A/\Lambda^A)/(\bar{\Lambda}^D/\Lambda^D)$ and
$({\bar{p}}^A/{p}^A)/({\bar{p}}^A/p^A)$ are
accessible in experiment and
the behaviors of
$(\bar{\Lambda}^A/\Lambda^A)/(\bar{\Lambda}^D/\Lambda^D)$ and
$({\bar{p}}^A/{p}^A)/({\bar{p}}^A/p^A)$ are different for different
models. According to Fig.~\ref{ratio} and Fig.~\ref{ratiop} , we
conclude that the various models about the EMC effect with different
sea behaviors can be distinguished by the future data of the
$x$-dependence of
$(\bar{\Lambda}^A/\Lambda^A)/(\bar{\Lambda}^D/\Lambda^D)$ and
$({\bar{p}}^A/{p}^A)/({\bar{p}}^A/p^A)$ in semi-inclusive deep
inelastic scattering process. The difference between the pion excess
model and cluster model is not good enough to be checked out in
experiment, but whether the sea quark is enhanced or not is clear to
be distinguished according to the result.


{\bf Acknowledgements}  This work is partially supported by National
Natural Science Foundation of China (Nos.~10421503, 10575003,
10528510), by the Key Grant Project of Chinese Ministry of Education
(No.~305001), and by the Research Fund for the Doctoral Program of
Higher Education (China).


\end{document}